\documentclass[preprint,12pt,compress]{elsarticle}
\usepackage{graphicx}
\usepackage{dcolumn}
\usepackage{bm}
\usepackage{amsmath}
\usepackage{latexsym}
\usepackage{epsfig}
\usepackage{amsbsy}
\usepackage{array}
\usepackage{setspace}
\usepackage{bm}
\usepackage{epstopdf}
\usepackage{slashed}
\usepackage{multirow}

\journal{Annals of Physics}

\begin{document}

\begin{frontmatter}
\title{Improve Microwave Quantum Illumination Via Optical Parametric
Amplifier}
\author{Biao Xiong, Xun Li, Xiao-Yu Wang, Ling Zhou\thanks{%
Corresponding author:\hspace{0.2cm}zhlhxn@dlut.edu.cn}}
\address{ School of Physics, Dalian
University of Technology, Dalian 116024, People's Republic of China}

\begin{abstract}
Quantum illumination is a quantum-optical sensing technique in which an
entangled source is exploited to improve the detection of a low-reflectivity
object that is immersed in a bright thermal background. Entangled sources
between microwave and optical fields can be exploited to improve detection
in microwave quantum illumination technique. We proposed a scheme to enhance the entanglement between the output
fields of microwave and optical cavity by introducing   optical parametric amplifier medium in cavity optomechanical system. We show that improving
signal-to-noise ratio and decreasing error probability of detection can be
obtained consequently even for objects with low reflectivity in the presence
of optical parametric amplifier.
\end{abstract}

\begin{keyword}
Microwave quantum illumination, Optical parametric amplifier,
Lower error probability

\end{keyword}

\end{frontmatter}

\section{Introduction}

The concept of quantum illumination was first put forward by S. Lloyd \cite%
{Lloyd1463}, where one of a pair of entangled photons is sent into a target
region, the other photon is retained at the receiver; a joint measurement
between the reflected back from target region and the retained photon can
discriminate whether the object presents or not.The quantum illumination is
of great significance. By the technique, we can defeat passive eavesdropping %
\cite{PhysRevA.80.022320}. Compared with traditional radar, quantum
illumination has the characteristics of higher signal-to-noise ratio and the
lower error probability \cite%
{PhysRevA.90.052308,PhysRevLett.114.110506,PhysRevA.89.062309} . The
single-photon quantum illumination \cite{Lloyd1463} was extended into
Gaussian-state quantum illumination very soon \cite%
{PhysRevLett.101.253601,PhysRevA.80.052310,1367-2630-11-6-063045}. Based on %
\cite{Lloyd1463} and \cite{PhysRevLett.101.253601} schemes, E. D. Lopaeva
et. al. experimentally realized the quantum illumination protocol at the
optical wave lengths \cite{PhysRevLett.110.153603}.

On the other hand, optomechanical interaction can couple high-frequency
optical field with low-frequency field with nonlinear interaction, which
offer us many amazing potential application such as entangling macroscopic
and microscopic object \cite%
{PhysRevLett.112.110406,PhysRevA.86.042323,PhysRevA.88.062341,PhysRevA.89.022335,PhysRevA.93.033842,PhysRevA.91.022326,PhysRevA.86.013809,PhysRevA.91.013807}
, performing high-precision measurements \cite%
{PhysRevLett.113.151102,Abbott2009Observation,PhysRevLett.114.113601},
processing quantum information \cite%
{PhysRevA.91.063836,doi:10.1080/09500340.2014.927016,0953-4075-48-3-035503}
and cooling a mechanical oscillator to its quantum ground state \cite%
{doi:10.1080/09500340802454971,PhysRevA.93.063853}, producing sideband comb %
\cite{Xiong201443}. The entanglement between optical field and microwave
field provides us the possibility to detect invisible object. Most recently,
S. Barzanjeh $et$ $al$. theoretically extended optical quantum illumination
to microwave quantum illumination by using an electro-optomechanical (EOM)
converter where the optomechanical system capacitively couples with LC
oscillating circuit through the movement of the mechanical oscillator driven
by the optical driven fields \cite{PhysRevLett.114.080503}. In Shabir
Barzanjeh's quantum-illumination protocol, microwaves produced by the LC
circuit and emitted to the target area can entangle the optical waves which
are retained to operate a jointed measurement with the reflected signal.
They showed that microwave quantum illumination dramatically outperforms a
conventional (coherent state) microwave radar of the same transmitted
energy, achieving an orders-of-magnitude lower detection-error probability,
which is related to the entanglement degree. In optomechanical system, there
are many ways to improve entanglement \cite%
{Xiao2013Entanglement,Cheng2015Preservation,Cheng2015590,Bai2016Robust,PhysRevA.94.012334}%
, one of which is to put an optical parametric amplifier (OPA) crystal in an
optomechanical system \cite{Pan2016}.

Inspired by the ideas elaborated in Refs.\cite%
{PhysRevLett.101.253601,PhysRevA.80.052310,1367-2630-11-6-063045,PhysRevLett.114.080503,Pan2016}%
, we put forward a scheme to improve the signal-to-noise ratio of microwave
quantum illumination by introducing $\chi ^{(2)}$ nonlinear medium. We show
that the entanglement between optical field and the microwave field can be
enhanced because of the enlarged effective optomechanical coupling strength,
meanwhile the signal-to-noise ratio is improved and the error probability of
detection can be decreased.

\begin{figure}[tbp]
\centering   \includegraphics[width=4.9in]{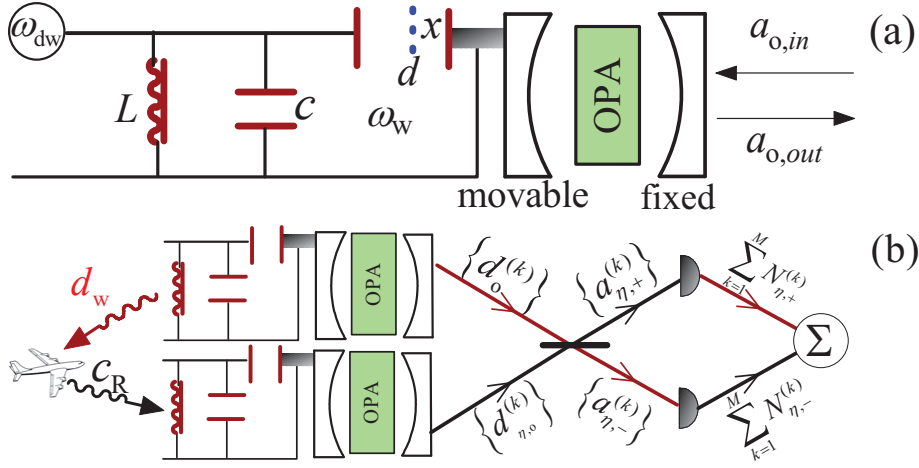}
\caption{(a): Schematic of the EOM system. The microwave cavity and the
optical cavity contained an OPA crystal can be coupled simultaneously with
mechanical oscillator. (b): Schematic of quantum illumination, which is
composed of two EOM system, one of which is used as the transmitter to
transmit the entangled microwave and optical idler fields, and the other as
a receiver, which receives the reflected microwave, and it's optical output
is used in a joint measurement with the retained idler. }
\label{fig:fig1}
\end{figure}

\section{Model and the Hamiltonian of the system}

The EOM system is shown in Fig.\ref{fig:fig1}(a), a mechanical resonator
(MR) capacitively couples to the microwave cavity (MC) consisting of LC
circuit with resonant frequency $\omega _{w}$ on the one side, and on the
other side couples to the optical cavity (OC) with resonant frequency $%
\omega _{o}$ in which lied a $\chi ^{(2)}$ nonlinear medium. The microwave
quantum illumination device as shown in Fig.\ref{fig:fig1}(b) contains two
identical EOM system, one of which is used as the transmitter to transmit
the microwaves and the idle optical waves, and the other as a receiver,
which receives the reflected microwaves, and can output optical waves \cite%
{PhysRevLett.114.080503}. For convenience, we define them as EOM
transmitters and EOM receivers, respectively. The Hamiltonian of the hybrid
system in the interaction picture reads \cite{PhysRevLett.114.080503,
PhysRevA.76.042336,PhysRevA.80.033807,PhysRevA.84.042342,PhysRevLett.109.130503,PhysRevLett.110.253601,PhysRevA.91.013807}
\begin{equation}
{H}=\hbar \omega _{m}{b}^{\dagger }{b}+\hbar \sum_{j=w,o}{\left[ \Delta
_{j}-g_{j}(b+b^{\dagger })\right] a_{j}^{\dagger }a_{j}}+i\hbar
E_{j}(a_{j}^{\dagger }-a_{j})+i\hbar G(e^{i\theta }a_{o}^{\dagger
2}-e^{-i\theta }a_{o}^{2}),
\end{equation}%
where $a_{j}$ ($b$) is the annihilation operator of cavity field (mechanical
resonator) with frequency $\omega _{j}$ $(\omega _{m})$. $\Delta _{j}=\omega
_{j}-\omega _{d,j}$ denotes the detuning of the two cavities frequencies $%
\omega _{j}$ from their driving field frequencies $\omega _{d,j}$, with $%
j=w,o$ describing the microwave cavity and optical cavity, respectively. $%
g_{o}=\frac{\omega _{o}}{l}\sqrt{\frac{\hbar }{2m\omega _{m}}}$ gives the
OC-MR coupling rate, with $m$ the effective mass of MR, and $l$ the length
of the optical cavity, while the MC-MR coupling $g_{w}=\frac{\mu \omega _{w}%
}{2d}\sqrt{\frac{\hbar }{2m\omega _{m}}}$ is corresponding to the effective
mass $m$ of MR and a separation $d$ of the two plates marked in Fig.\ref%
{fig:fig1}(a), and $\mu =C/(C+C_{0})$ is the dimensionless parameter, with $%
C $ being the capacitance of parallel capacitor, and $C_{0}$ being the bare
capacitance corresponding to the $d$ separation of the two palates. The
driving strength of cavity $E_{j}=\sqrt{2P_{j}\kappa _{j}/\hbar \omega _{d,j}%
}$ with the power $P_{j}$. And $\kappa _{j}$ describes the losing rates of
cavity $j$. $G$ is the nonlinear gain of the OPA, and $\theta $ is the phase
of the field driving the OPA. Considering the loss rates of the cavities and
resonator, we can give the nonlinear Heisenberg-Langevin equation 
\begin{eqnarray}
\dot{a}_{w} &=&(-i\Delta _{w}-\kappa _{w})a_{w}+ig_{w}(b+b^{\dagger
})a_{w}+E_{w}+\sqrt{2\kappa _{w}}a_{w,in},  \label{Nlangevin1} \\
\dot{a}_{o} &=&(-i\Delta _{o}-\kappa _{o})a_{o}+ig_{o}(b+b^{\dagger
})a_{o}+E_{o}+2Ge^{i\theta }a_{o}^{\dagger }+\sqrt{2\kappa _{o}}a_{o,in},
\label{Nlangevin2} \\
\dot{b} &=&(-i\omega _{m}-\gamma _{m})b+ig_{o}a_{o}^{\dagger
}a_{o}+ig_{w}a_{w}^{\dagger }a_{w}+\sqrt{2\gamma _{m}}b_{in},
\label{Nlangevin3}
\end{eqnarray}%
where $a_{w,in}$, $a_{o,in}$ and $b_{in}$ are the input noise operators of
the microwave, of the optical and of the mechanical resonator, with the
following correlation functions 
\begin{eqnarray}
\langle a_{j,in}(t)a_{j,in}^{\dagger }(t^{\prime })\rangle &=&(\bar{n}%
_{j}^{T}+1)\delta (t-t^{\prime }),  \label{correlation1} \\
\langle b_{in}(t)b_{in}^{\dagger }(t^{\prime })\rangle &=&(\bar{n}%
_{b}^{T}+1)\delta (t-t^{\prime }),  \label{correlation2}
\end{eqnarray}%
where $\bar{n}_{j}^{T}=[exp(\hbar \omega _{j}/k_{B}T)-1]^{-1}$ is the
equilibrium mean thermal photon numbers of the optical ($j=o$) and microwave
($j=w$) fields. And $\bar{n}_{b}^{T}=[exp(\hbar \omega _{m}/k_{B}T)-1]^{-1}$
is the mean thermal excitation numbers of the resonator. We can safely
assume $\bar{n}_{o}^{T}\approx 0$ since $\hbar \omega _{j}/k_{B}T\gg 1$ at
optical frequencies and low temperatures, while thermal microwave photons
and thermal excitation phonons of resonator cannot be neglected usually,
even at very low temperatures.

\begin{figure}[tbp]
\centering   \includegraphics[width=4.2in]{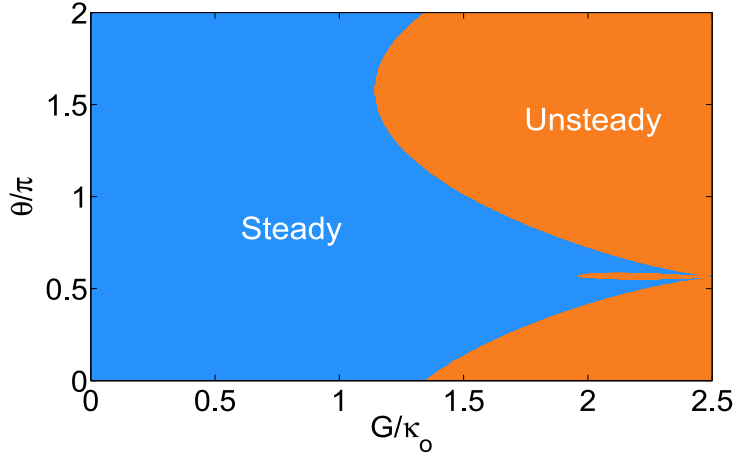}
\caption{The stability of the system affected by $G/\protect\kappa _{o}$ and 
$\protect\theta /2\protect\pi $ , where $T=30$ mk, $\protect\lambda =1064$
nm, $\protect\omega _{m}/2\protect\pi =10$ MHz, $Q=\protect\omega _{m}/%
\protect\gamma _{m}=30\times 10^{3}$, $\protect\omega _{w}/2\protect\pi =10$
GHz, $g_{\mathrm{w}}/2\protect\pi =0.327$, $g_{\mathrm{o}}/2\protect\pi %
=115.512$, $\Delta _{\mathrm{w}}=-\protect\omega _{m},\Delta _{\mathrm{o}}=%
\protect\omega _{m}$, $\protect\kappa _{\mathrm{w}}=0.24\protect\omega _{m},%
\protect\kappa _{\mathrm{o}}=0.2\protect\omega _{m}$, $P_{\mathrm{o}}=10P_{%
\mathrm{w}}=10$ mW. }
\label{fig:fig2new}
\end{figure}

We can expand the operator of the optical field as its steady-state mean
value and a small fluctuation with zero mean value under the strong driving
condition, i.e. $a_{j}\rightarrow \alpha _{j}+a_{j}$. By setting the
derivatives of $\alpha _{j}$ to zero, we obtain the steady-state mean values 
$\alpha _{j}$ as 
\begin{eqnarray}
\alpha _{w} &=&\frac{E_{w}}{i\Delta _{w}+\kappa _{w}},  \label{steady1} \\
\alpha _{o} &=&\frac{(-i\Delta _{o}+\kappa _{o}+2Ge^{i\theta })E_{o}}{\kappa
_{o}^{2}+\Delta _{o}^{2}-4G^{2}}.  \label{steady2}
\end{eqnarray}%
Accordingly, the linearized Langevin equations for fluctuation operators
become 
\begin{eqnarray}
\dot{a}_{w} &=&(-i\Delta _{w}-\kappa _{w})a_{w}+ig_{w}^{\prime
}(b+b^{\dagger })+\sqrt{2\kappa _{w}}a_{w,in},  \label{linear1} \\
\dot{a}_{o} &=&(-i\Delta _{o}-\kappa _{o})a_{o}+ig_{o}^{\prime
}(b+b^{\dagger })+2Ge^{i\theta }a_{o}^{\dagger }+\sqrt{2\kappa _{o}}a_{o,in},
\label{linear2} \\
\dot{b} &=&(-i\omega _{m}-\gamma _{m})b+ig_{o}^{\prime }a_{o}^{\dagger
}+ig_{o}^{\prime \ast }a_{o}+ig_{w}^{\prime }a_{w}^{\dagger }+ig_{w}^{\prime
\ast }a_{w}+\sqrt{2\gamma _{m}}b_{in},  \label{linear3}
\end{eqnarray}%
where $g_{j}^{\prime }=\alpha _{j}g_{j}$ ($j=w,o$) is the effective coupling
rate. We rewrite Eqs.~(\ref{linear1})-(\ref{linear3}) and the corresponding
transposed conjugate in the compact form 
\begin{eqnarray}\label{compact}
\dot{f}=\Lambda f+\zeta, 
\end{eqnarray}%
where $f=(a_{w},a_{w}^{\dagger },a_{o},a_{o}^{\dagger },b,b^{\dagger })^{T}$%
, $\zeta =(\sqrt{2\kappa _{w}}a_{w,in},\sqrt{2\kappa _{w}}a_{w,in}^{\dagger
},\sqrt{2\kappa _{o}}a_{o,in},\newline
\sqrt{2\kappa _{o}}a_{o,in}^{\dagger },\sqrt{2\gamma _{m}}b_{in},\sqrt{%
2\gamma _{m}}b_{in}^{\dagger })^{T}$ where the superscript $T$ means
transposition, and the coefficient matrix $\Lambda $ is given by {\small 
\begin{eqnarray}\label{matrix}
\left[ 
\begin{array}{cccccc}
-i\Delta _{w}-\kappa _{w} & 0 & 0 & 0 & ig_{w}^{\prime } & ig_{w}^{\prime }
\\ 
0 & i\Delta _{w}-\kappa _{w} & 0 & 0 & -ig_{w}^{\prime \star } & 
-ig_{w}^{\prime \star } \\ 
0 & 0 & -i\Delta _{o}-\kappa _{o} & 2Ge^{i\theta } & ig_{o}^{\prime } & 
ig_{o}^{\prime } \\ 
0 & 0 & 2Ge^{-i\theta } & i\Delta _{o}-\kappa _{o} & -ig_{o}^{\prime \star }
& -ig_{o}^{\prime \star } \\ 
ig_{w}^{\prime \star } & ig_{w}^{\prime } & ig_{o}^{\prime \star } & 
ig_{o}^{\prime } & -i\omega _{m}-\gamma _{m} & 0 \\ 
-ig_{w}^{\prime \star } & -ig_{w}^{\prime } & -ig_{o}^{\prime \star } & 
-ig_{o}^{\prime } & 0 & i\omega _{m}-\gamma _{m}%
\end{array}%
\right].
\end{eqnarray}%
}The system is stable only if all of the eigenvalues of $\Lambda $ are of
real negative parts, which means Routh-Hurwitz criterion \cite%
{PhysRevA.35.5288} satisfied. Because of the matrix $\Lambda $ is $6\times 6$
dimensions, so it is difficult to obtain the analytical solution of steady
condition. We assume that $\Delta _{w}=-\Delta _{o}=-\omega _{m}$, i.e., the
optical field satisfies blue-sideband condition while the microwave field is
in the red-sideband. We numerically solve the eigenvalues of $\Lambda $ to
ensure the stability of our system, which is shown in Fig.~\ref{fig:fig2new}.
It is clearly shows that the system can reach its steady state at a wide
range of $G$ and $\theta $, although the system contains a nonlinear
medium with amplification effect. Hereafter, we will choose parameters
within steady region.

We can easily solve Eqs.~(\ref{linear1})-(\ref{linear3}) in the frequency
domain by Fourier transform $O(\omega )=\frac{1}{\sqrt{2\pi }}\int
dtO(t)e^{i\omega t}$. By substituting the solutions of Eqs.~(\ref{linear1})-(%
\ref{linear3}) in frequency domain into the standard input-output relation $%
d_{j}=a_{j,out}=\sqrt{2\kappa }a_{j}-a_{j,in}$, we have 
\begin{eqnarray}
d_{w}(\omega ) &=&A_{1}(\omega )a_{w,in}(\omega )+A_{2}(\omega
)a_{w,in}^{\dagger }(\omega )+A_{3}(\omega )a_{o,in}(\omega )  \notag \\
&&+A_{4}(\omega )a_{o,in}^{\dagger }(\omega )+A_{5}(\omega )b_{in}(\omega
)+A_{6}(\omega )b_{in}^{\dagger }(\omega ),  \label{output1} \\
d_{o}(\omega ) &=&B_{1}(\omega )a_{w,in}(\omega )+B_{2}(\omega
)a_{w,in}^{\dagger }(\omega )+B_{3}(\omega )a_{o,in}(\omega )  \notag \\
&&+B_{4}(\omega )a_{o,in}^{\dagger }(\omega )+B_{5}(\omega )b_{in}(\omega
)+B_{6}(\omega )b_{in}^{\dagger }(\omega ),  \label{output2}
\end{eqnarray}%
where the coefficients $A_{j}(\omega )$ and $B_{j}(\omega )$ ($j=1,2...6$)
are given in \ref{The expression of output fields} (\ref{A1}-\ref{A19}). We
see clearly that $d_{w}(\omega )$ is contributed partly by optical field
(contain $a_{o,in}(\omega )$ and $a_{o,in}^{\dagger }(\omega )$), similarly, 
$d_{o}(\omega )$ is contributed partly by microwave field (contain $%
a_{w,in}(\omega )$ and $a_{w,in}^{\dagger }(\omega )$), which exhibits that
there is quantum correlation between the optical field and microwave field.
The correlation functions of Eq.~(\ref{correlation1}) and (\ref{correlation2}%
) are transformed to the frequency domain as 
\begin{eqnarray}
\langle a_{j,in}(\omega )a_{j,in}^{\dagger }(\omega ^{\prime })\rangle &=&(%
\bar{n}_{j}^{T}+1)\delta (\omega +\omega ^{\prime }),  \label{fcorrelation1}
\\
\langle b_{in}(\omega )b_{in}^{\dagger }(\omega ^{\prime })\rangle &=&(\bar{n%
}_{b}^{T}+1)\delta (\omega +\omega ^{\prime }).  \label{fcorrelation2}
\end{eqnarray}%

We define $n(o|w)$ to describe the mean number of microwave fields
contributed by the output optical fields. Similarly, $n(w|o)$ describes the
mean number of photons contributed by the output microwave fields. From Eqs.~(\ref{output1})
 and ~(\ref{output2}), we have 
\begin{eqnarray}
n(o|w) &=&(|B_{1}(\omega )|^{2}+|B_{2}(\omega )|^{2})\bar{n}%
_{w}^{T}+|B_{2}(\omega )|^{2},  \label{WintoO} \\
n(w|o) &=&(|A_{3}(\omega )|^{2}+|A_{4}(\omega )|^{2})\bar{n}%
_{o}^{T}+|A_{4}(\omega )|^{2}.  \label{OintoW}
\end{eqnarray}

\section{MICROWAVE-OPTICAL OUTPUT ENTANGLEMENT}

Now we introduce the covariance matrix (CM) $V$ in the frequency domain. The
matrix element of the $4\times 4$ CM can be expressed as 
\begin{equation}
\delta (\omega +\omega ^{\prime })V_{ij}(\omega )=\frac{1}{2}\langle
u_{i}(\omega )u_{j}(\omega ^{\prime })+u_{j}(\omega ^{\prime })u_{i}(\omega
)\rangle ,  \label{corva1}
\end{equation}%
where 
\begin{equation}
\mathbf{u}(\omega )=[X_{w}(\omega ),Y_{w}(\omega ),X_{o}(\omega
),Y_{o}(\omega )]^{T},  \label{corva2}
\end{equation}%
and $X_{j}(\omega )=(d_{j}(\omega )+d_{j}^{\dagger }(\omega ))/\sqrt{2}$, $%
Y_{j}(\omega )=(d_{j}(\omega )-d_{j}^{\dagger }(\omega ))/i\sqrt{2}$ with $%
j=o,w$ denoting the quadratures fluctuation of the optical and microwave
field, respectively.

With the help of CM, we can easily quantify the output entanglement of the
microwave fields and optical fields using logarithmic negativity \cite{PhysRevA.70.022318}, i.e., 
\begin{equation}
E_{N}=\mathrm{max}\{0,-\mathrm{ln}2\eta ^{-}\},
\end{equation}%
where $\eta ^{-}=2^{-1/2}\left[ \Sigma (V)-\sqrt{\Sigma (V)^{2}-4\mathrm{det}%
V}\right] ^{1/2}$ is the lowest symplectic eigenvalue of the partial
transpose of the CM, and $\Sigma (V)=\mathrm{det}A+\mathrm{det}B-2\mathrm{det%
}C$, with the $A$, $B$, $C$ being the $2\times 2$ matrix, taking from the CM 
\begin{equation}
V=\left( 
\begin{array}{cc}
A & C \\ 
C^{T} & B%
\end{array}%
\right) .
\end{equation}

\begin{figure}[tbp]
\centering   \includegraphics[width=5.1in]{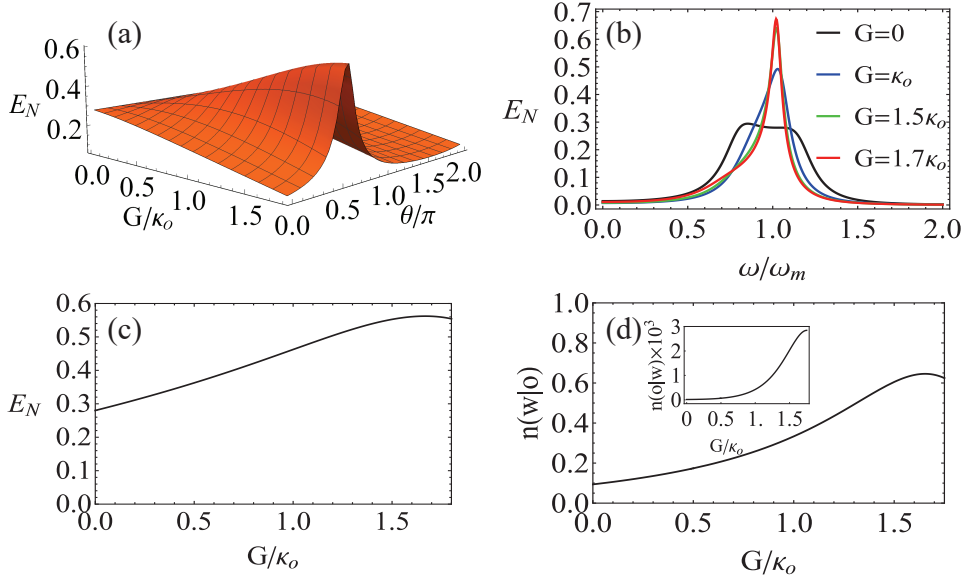}
\caption{(a) $E_{N}$ versus $G/\protect\kappa _{o}$ and $\protect\theta /2%
\protect\pi $, (b) $E_{N}$ versus $\protect\omega /\protect\omega _{m}$, (c) 
$E_{N}$ versus $G/\protect\kappa _{o}$, and (d) $n(o|w)$ ($n(w|o)$ in
subgraph) versus $G/\protect\kappa _{o}$. We set $\protect\omega =\protect%
\omega _{m}$ for (a),(c),(d) and $\protect\theta =0.62\protect\pi $ for
(b),(c),(d). The other parameters are the same as in Fig.\ref{fig:fig2new}}
\label{fig:fig3new}
\end{figure}

We plot the output entanglement between the cavity field and microwave field
in Fig.3. \ We also keep all of the parameters meeting the steady state
condition that displayed in Fig.\ref{fig:fig2new}. As shown in Fig.\ref%
{fig:fig3new}(a), the logarithmic negativity $E_{N}$ is increased with the
increasing of the value of the nonlinear gain $G$ around $\theta =0.62\pi $.
Fig.\ref{fig:fig3new}(b) also exhibits that the maximum value of $E_{N}$
with nonzero value of $G$ is obviously larger than that with $G=0$,
meanwhile, with the increasing of $G$, the maximum value of entanglement is
enhanced. From the definition of $n(o|w)$ and $n(w|o)$, we know that they
exhibit quantum correlation and should have relation with entanglement. Eq.~(%
\ref{WintoO}) and (\ref{OintoW}) is visualized in Fig.\ref{fig:fig3new} (d)
and its subgraph. We see that $n(o|w)$ and $n(w|o)$ increase with the
increasing of $G$ which is consistent with Fig.\ref{fig:fig3new}(c).

In order to understand the effect of OPA medium on the entanglement between
the cavity field and microwave field, we perform a squeezing transformation %
\cite{PhysRevLett.114.093602,Wang2016,Wang20162} 
\begin{eqnarray}
a_{o}=cosh(r)\widetilde{a}_{o}-ie^{i\theta }sinh(r)\widetilde{a}%
_{o}^{\dagger },
\end{eqnarray}%
where $r=\frac{1}{4}ln\frac{\Delta _{o}+2G}{\Delta _{o}-2G}$. Then the
Hamiltonian can be written as 
\begin{eqnarray}
{H} &=&\hbar \omega _{m}{b}^{\dagger }{b}+\hbar \Delta _{o}^{\prime }%
\widetilde{a}_{o}^{\dagger }\widetilde{a}_{o}-\hbar g_{os}\widetilde{a}%
_{o}^{\dagger }\widetilde{a}_{o}(b+b^{\dagger })+\hbar g_{op}(e^{-i(\theta +%
\frac{\pi }{2})}\widetilde{a}_{o}^{2}+e^{i(\theta +\frac{\pi }{2})}%
\widetilde{a}_{o}^{\dagger 2})  \notag \\
&&(b+b^{\dagger })+\hbar \Delta _{w}a_{w}^{\dagger }a_{w}-\hbar
g_{w}a_{w}^{\dagger }a_{w}(b+b^{\dagger })+H_{dri}^{\prime },
\label{squeezingH}
\end{eqnarray}%
with 
\begin{equation*}
H_{dri}^{\prime }=i\hbar E_{w}(a_{w}^{\dagger }-a_{w})+i\hbar
E_{o}[(cosh(r)+e^{i(\theta +\frac{\pi }{2})}sinh(r))\widetilde{a}%
_{o}^{\dagger }-h.c.],
\end{equation*}%
where $\Delta _{O}^{\prime }=\Delta _{o}cosh(2r)-2Gsinh(2r)$, $g_{os}=\frac{%
g_{o}\Delta _{o}}{\sqrt{\Delta _{o}^{2}-4G^{2}}}$, $g_{op}=\frac{g_{o}G}{%
\sqrt{\Delta _{o}^{2}-4G^{2}}}$. We can see that the third and fourth terms
in Eq. (\ref{squeezingH}) describe the standard optomechanical
radiation-pressure and parametric amplification interactions, respectively,
with the controllable strengths $g_{os}$ and $g_{op}$. The strength of the
optomechanical radiation-pressure $g_{os}$ is enhanced than $g_{o}$, i.e.,$%
g_{os}\succ g_{o}$. The entanglement between the cavity field and microwave
field is resulted from the optomechanical radiation-pressure interactions
via a common bus of mechanical oscillator; therefore, the entanglement can
be enhanced by the participating of $\chi ^{(2)}$ medium. The forth term
also should contribute to the squeezing of the optical mode.

\section{ERROR PROBABILITY FOR DETECTION}

As shown in Fig.1b, the microwave emitted by the EOM transmitter is shined
on the surface of the target object, the reflected back signal by the object
is the input signal to the EOM receiver. The optical output of the receiver,
similar with Eq.(\ref{output2}), can be written as 
\begin{eqnarray}\label{output} 
d_{\eta ,o}(\omega ) &=&B_{1}(\omega )\,a_{R}(\omega )+B_{2}(\omega
)\,a_{R}^{\dagger }(\omega )+B_{3}(\omega )\,a_{o,in}(\omega )  \notag \\
&&+B_{4}(\omega )\,a_{o,in}^{\dagger }(\omega )+B_{5}(\omega
)\,b_{in}(\omega )+B_{6}(\omega )\,b_{in}^{\dagger }(\omega ),
\end{eqnarray}%
where we have defined a new input operator of microwave
cavity $a_{R}=a_{B}$ under the hypothesis that the target region do not
contain a low-reflectivity object (hypothesis $H_{0}$), and $a_{R}=\sqrt{%
\eta }\,d_{w}+\sqrt{1-\eta }\,a_{B}$ under the hypothesis that the target
region contain a low-reflectivity object (hypothesis $H_{1}$). Here $a_{B}$
is the annihilation operator of the back ground noise which is in a thermal
state with the photon number $n_{B}$ with hypothesis $H_{0}$ and $%
n_{B}/(1-\eta )$ with hypothesis $H_{1}$. We can safely assume $\langle
a_{B}^{\dagger }a_{B}\rangle =n_{B}$ for both hypothesis, since the
effective reflectivity (including propagation losses and target
reflectivity) $\eta $ is small and $n_{B}$ is very large.

\begin{figure}[]
\centering   \includegraphics[width=5.2in]{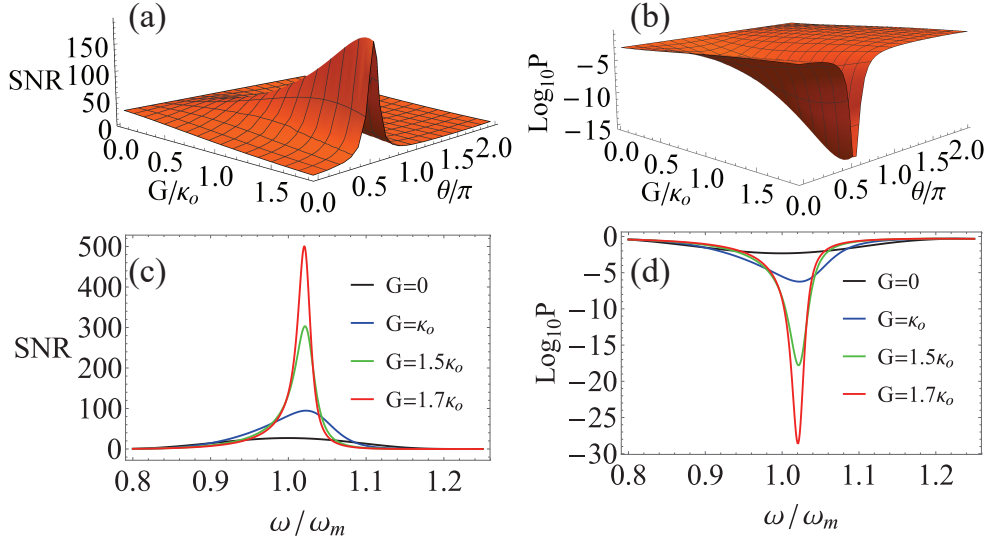}
\caption{Plot of: (a) signal-to-noise ratio $\mathrm{SNR}$ and (b) error
probability $\mathrm{P}$ as a function of $G$ and $\protect\theta$. (c)
signal-to-noise ratio $\mathrm{SNR}$ and (d) error probability $\mathrm{P}$
versus $\protect\omega/\protect\omega_{m}$. We set $\protect\eta=0.07$, $%
M=10^6$, $n_{B}=610$ and the other parameters are the same as in Fig. \ref%
{fig:fig2new}. }
\label{fig:fig4new}
\end{figure}

As shown in Fig.\ref{fig:fig1}(b), the returned optical signal is combined
with the retained idler on a 50-50 beam splitter whose output is 
\begin{equation}
a_{\eta ,\pm }=\frac{d_{\eta ,o}\pm d_{o}}{\sqrt{2}}.
\end{equation}%

For quantum illumination target detection, our signal-idler mode pair
analysis must be extended to a continuous-wave EOM system whose $W_{m}$%
-Hz-bandwidth output fields are used in a $t_{m}$-sec-duration measurement
involving $M=t_{m}W_{m}\gg 1$ independent, identically-distributed mode
pairs to discriminate target absence from target presence \cite%
{PhysRevLett.111.010501}. The identically-distribute mode pairs are then
photo-detected. Finally, the target absence-or-presence decision is made by
comparing the difference of the two detectors' total photon counts 
\begin{equation}
N_{\eta }=\sum_{k=1}^{M}\left( N_{\eta ,+}^{(k)}-N_{\eta ,-}^{(k)}\right),
\end{equation}%
where $N_{\eta ,\pm }^{(k)}=a_{\eta ,\pm }^{(k)\dagger }\,a_{\eta ,\pm
}^{(k)}$ is corresponding to the photon-counts. The error probability of
quantum illumination can be expressed as \cite{PhysRevLett.114.080503} 
\begin{equation}
\mathrm{P}=\frac{{\mathrm{erfc}}\left( \mathrm{SNR}/8\right) }{2},  \label{P}
\end{equation}%
where the $M$-Mode signal-to-noise ratio is 
\begin{equation}
\mathrm{SNR}=\frac{4M[(\langle N_{\eta ,+}\rangle _{H_{1}}-\langle N_{\eta
,-}\rangle _{H_{1}})-(\langle N_{\eta ,+}\rangle _{H_{0}}-\langle N_{\eta
,-}\rangle _{H_{0}})]^{2}}{\left( \sqrt{\langle (\Delta N_{\eta ,+}-\Delta
N_{\eta ,-})^{2}\rangle _{H_{0}}}+\sqrt{\langle (\Delta N_{\eta ,+}-\Delta
N_{\eta ,-})^{2}\rangle _{H_{1}}}\right) ^{2}}.  \label{SNR}
\end{equation}%
Eqs~.(\ref{P}) (\ref{SNR}) can be calculated easily, since we have calculated $%
\langle N_{\eta }\rangle _{H_{j}}$ and $\langle (\Delta N_{\eta ,+}-\Delta
N_{\eta ,-})^{2}\rangle _{H_{j}}$ in \ref{The calculation of Signal-to-noise}
(\ref{photoncounts1}-\ref{photoncounts2}).

Now we discuss the properties of quantum illumination. We numerically
simulate the signal-to-noise ratio $\mathrm{SNR}$ and error probability $%
\mathrm{P}$ as a function of $G$ and $\theta $ or $\omega /\omega _{m}$,
which is shown in Fig.\ref{fig:fig4new}(a), Fig.\ref{fig:fig4new}(b), Fig.%
\ref{fig:fig4new}(c) and Fig.\ref{fig:fig4new}(d) respectively.The maximum
value of signal-to-noise ratio $\mathrm{SNR}$ with OPA is higher than that
without OPA [see Figs.4(a) and 4(c)], and the error probability $\mathrm{P}$
with OPA is decreased comparing with that of without OPA. Moreover, the
larger values of $G$, the higher value of SNR and lower value of error
probability $\mathrm{P.}$ Jointly considering Fig.\ref{fig:fig4new} with Fig.%
\ref{fig:fig3new}, we see that the larger the entanglement, the higher the
signal-to-noise ratio $\mathrm{SNR}$, and the lower the error probability of
detection. That is to say, when we can archive ideal entanglement in the
region in terms of the parameters $G$ , $\theta $ and $\omega /\omega _{m}$,
the signal-to-noise ratio $\mathrm{SNR}$ also obtain its maximum values with
the minimum error probability $\mathrm{P.}$

\begin{figure}[tbp]
\centering   \includegraphics[width=5.2in]{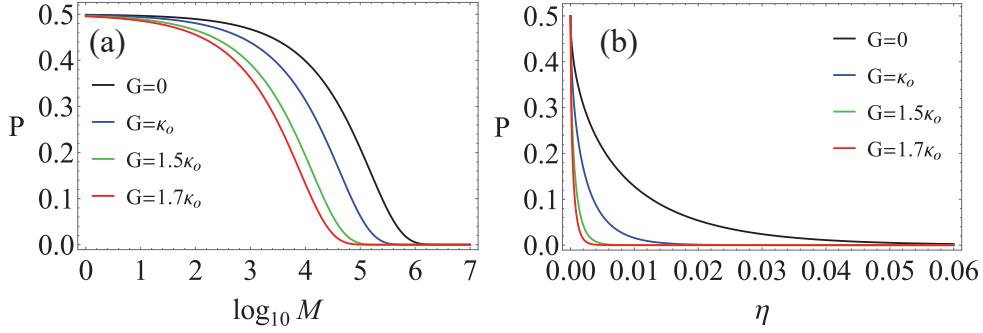}
\caption{Plot of error probability $\mathrm{P}$ versus (a) mode pairs $M$
and (b) efficient reflectivity $\protect\eta $. We set $\protect\eta =0.05$
for (a), $M=10^{6}$ for (b). $%
\protect\theta =0.62\protect\pi $ and $\protect\omega =1.02\protect\omega %
_{m}$ , $n_{B}=610$ for both . The other parameters see Fig. \ref%
{fig:fig2new}. }
\label{fig:fig5new}
\end{figure}

In order to further study the advantages of our microwave quantum
illumination system, we plot error probability $\mathrm{P}$ versus
identically-distributed mode pairs $M$ or efficient reflectivity rate $\eta $
at different OPA gain coefficient $G$, where we set $\eta =0.05$ for Fig.\ref%
{fig:fig5new}(a) and $M=10^{6}$ for Fig.\ref{fig:fig5new}(b).
From the simulation results shown in Fig.\ref{fig:fig5new}, we can see that
the superiority is obvious (lower error probability) in the presence of OPA.
Fig.\ref{fig:fig5new}(a) shows that for certain value of $M$, the error
probability $\mathrm{P}$ decreases with the increasing of values $G$, and
the more identically-distributed mode pairs $M$, the lower error probability 
$\mathrm{P}$. Furthermore, $P$ can reach almost zero when $M=10^{6}$ at $%
G=1.7\kappa _{o}$. Fig.\ref{fig:fig5new}(b) shows that without the OPA
medium, the error probability $\mathrm{P}$ decreases with the increase of
the efficient reflectivity $\eta $, but it still keep higher value (see the
black-line). With the help of OPA medium, the error probability $\mathrm{P}$
can be decreased greatly. Even for the low efficient reflectivity object,
the error probability $P$ still can achieve very low value with the
assistance of OPA medium, which is of great importance for practical
applications.

\begin{figure}[tbp]
\centering   \includegraphics[width=5.2in]{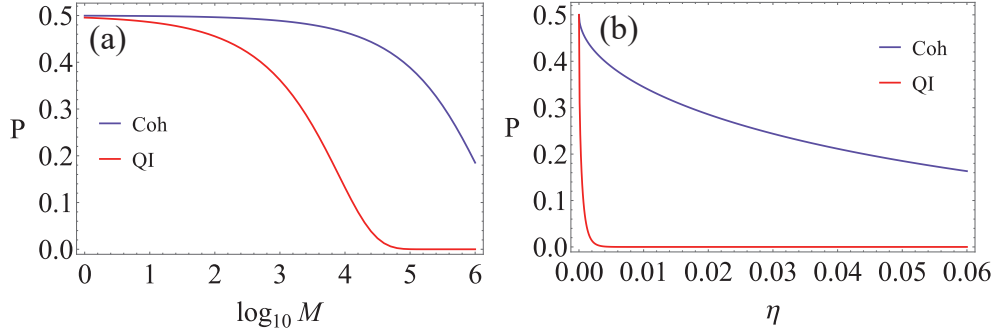}
\caption{The error probability $\mathrm{P}$ versus (a) mode pairs $M$ and
(b) efficient reflectivity $\protect\eta $. We set $\protect\eta =0.05$ for
(a), $M=10^{6}$ for (b). For both of (a) and (b), $\protect\theta =0.62%
\protect\pi $, $\protect\omega =1.02\protect\omega _{m}$, and $n_{B}=610$.
The other parameters are the same with Fig. \ref{fig:fig3new}. }
\label{fig:fig6new}
\end{figure}

We compare our quantum illumination system with the conventional radar,
whose error probability \cite{PhysRevA.80.052310} is 
\begin{eqnarray}
\mathrm{P_{coh}}=\frac{{\mathrm{erfc}}\left( \mathrm{SNR_{coh}}/8\right) }{2},
\end{eqnarray}%
where 
\begin{eqnarray}
\mathrm{SNR_{coh}}=\frac{4\eta Mn_{w}}{2n_{B}+1},
\end{eqnarray}%
here $n_{w}=\langle d_{w}^{\dagger }d_{w}\rangle =(|A_{1}(\omega
)|^{2}+|A_{2}(\omega )|^{2})\bar{n}_{w}^{T}+(|A_{3}(\omega
)|^{2}+|A_{4}(\omega )|^{2})\bar{n}_{o}^{T}+(|A_{5}(\omega
)|^{2}+|A_{6}(\omega )|^{2})\bar{n}_{b}^{T}+|A_{2}(\omega
)|^{2}+|A_{4}(\omega )|^{2}+|A_{6}(\omega )|^{2}$. We plot the error
probability of our quantum illumination and conventional radar as a function
of $M$ and $\eta $ in Fig.~\ref{fig:fig6new}(a) and Fig.~\ref{fig:fig6new}%
(b), which shows that our quantum illumination is distinct superior to
traditional radar. Here we set $G=1.7\kappa _{o}$.

\section{Summary}

In this paper, by introducing OPA medium, we show that the entanglement
between optical field and microwave field can be improved. We analyse the
mechanism that the OPA nonlinear medium affects the quantum entanglement,
meanwhile our investigation exhibits that the signal-to-noise ratio of the
detected object with low reflectivity rate $\eta $ can be improved, and the
the error probability of the detection can be reduced by introducing the OPA
nonlinear medium. That may give us effective way to improve microwave
quantum illumination. Since the OPA nonlinear medium has been widely used in
experiment and is mature technique, the presented scheme should be
applicable.

\section*{ACKNOWLEDGEMENT}

We would like to thank Mr.~Wen-Zhao Zhang and Jiong Cheng for helpful
discussions. This work was supported by the NSF of China under Grant numbers
11474044. \appendix

\section{The expression of output fields}

\label{The expression of output fields} The coefficients $A_{j}(\omega )$
and $B_{j}(\omega )$ of Eqs. (\ref{output1}),~(\ref{output2}) can be
expressed as {\small 
\begin{eqnarray}
A_{1}(\omega ) &=&\frac{1}{u}\frac{4i\omega _{m}|g_{w}^{\prime }|^{2}\kappa
_{w}}{\Delta _{w}^{-2}}-\frac{2\kappa _{w}}{\Delta _{w}^{-}}-1,  \label{A1} \\
A_{2}(\omega ) &=&\frac{1}{u}\frac{4i\omega _{m}g_{w}^{\prime 2}\kappa _{w}}{%
\Delta _{w}^{-}\Delta _{w}^{+}},  \label{A2} \\
A_{3}(\omega ) &=&-\frac{1}{u}\frac{4i\omega _{m}\sqrt{\kappa _{w}\kappa _{o}%
}(2Ge^{-i\theta }g_{o}^{\prime }g_{w}^{\prime }-\Delta _{o}^{+}g_{o}^{\prime
\ast }g_{w}^{\prime })}{\Delta _{w}^{-}(\Delta _{o}^{-}\Delta
_{o}^{+}-4G^{2})},  \label{A3} \\
A_{4}(\omega ) &=&-\frac{1}{u}\frac{4i\omega _{m}\sqrt{\kappa _{w}\kappa _{o}%
}(2Ge^{i\theta }g_{o}^{\prime \ast }g_{w}^{\prime }-\Delta
_{o}^{-}g_{o}^{\prime }g_{w}^{\prime })}{\Delta _{w}^{-}(\Delta
_{o}^{-}\Delta _{o}^{+}-4G^{2})},  \label{A4} \\
A_{5}(\omega ) &=&\frac{1}{u}\frac{2ig_{w}^{\prime }\sqrt{\kappa _{w}\gamma
_{m}}\Delta _{m}^{+}}{\Delta _{w}^{-}},  \label{A5} \\
A_{6}(\omega ) &=&\frac{1}{u}\frac{2ig_{w}^{\prime }\sqrt{\kappa _{w}\gamma
_{m}}\Delta _{m}^{-}}{\Delta _{w}^{-}}, \label{A6} \\
B_{1}(\omega ) &=&\frac{1}{u}\frac{4i\omega _{m}\sqrt{\kappa _{o}\kappa _{w}}%
(2Ge^{i\theta }g_{o}^{\prime \ast }g_{w}^{\prime \ast }+\Delta
_{o}^{+}g_{o}^{\prime }g_{w}^{\prime \ast })}{\Delta _{w}^{-}(\Delta
_{o}^{-}\Delta _{o}^{+}-4G^{2})},  \label{A7}\\
B_{2}(\omega ) &=&\frac{1}{u}\frac{4i\omega _{m}\sqrt{\kappa _{o}\kappa _{w}}%
(2Ge^{i\theta }g_{o}^{\prime \ast }g_{w}^{\prime }+\Delta
_{o}^{+}g_{o}^{\prime }g_{w}^{\prime })}{\Delta _{w}^{+}(\Delta
_{o}^{-}\Delta _{o}^{+}-4G^{2})},  \label{A8} \\
B_{3}(\omega ) &=&-\frac{1}{u}\frac{4i\omega _{m}\kappa _{o}(2Ge^{i\theta
}g_{o}^{\prime \ast }+\Delta _{o}^{+}g_{o}^{\prime })(2Ge^{-i\theta
}g_{o}^{\prime }-\Delta _{o}^{+}g_{o}^{\prime \ast })}{(\Delta
_{o}^{-}\Delta _{o}^{+}-4G^{2})^{2}}  \notag \\
&&-\frac{2\kappa _{o}\Delta _{o}^{+}}{\Delta _{o}^{-}\Delta _{o}^{+}-4G^{2}}%
-1 , \label{A9} \\
B_{4}(\omega ) &=&-\frac{1}{u}\frac{4i\omega _{m}\kappa _{o}(2Ge^{\mathrm{i}%
\theta }g_{o}^{\prime \ast }+\Delta _{o}^{+}g_{o}^{\prime })(2Ge^{\mathrm{i}%
\theta }g_{o}^{\prime \ast }-\Delta _{o}^{-}g_{o}^{\prime })}{(\Delta
_{o}^{-}\Delta _{o}^{+}-4G^{2})^{2}}  \notag \\
&&+\frac{4\kappa _{o}Ge^{\mathrm{i}\theta }}{\Delta _{o}^{-}\Delta
_{o}^{+}-4G^{2}},  \label{A10} \\
B_{5}(\omega ) &=&\frac{1}{u}\frac{2i\sqrt{\kappa _{o}\gamma _{m}}\Delta
_{m}^{+}(2Ge^{\mathrm{i}\theta }g_{o}^{\prime \ast }+\Delta
_{o}^{+}g_{o}^{\prime })}{\Delta _{o}^{-}\Delta _{o}^{+}-4G^{2}},  \label{A11}
\\
B_{6}(\omega ) &=&\frac{1}{u}\frac{2i\sqrt{\kappa _{o}\gamma _{m}}\Delta
_{m}^{-}(2Ge^{\mathrm{i}\theta }g_{o}^{\prime \ast }+\Delta
_{o}^{+}g_{o}^{\prime })}{\Delta _{o}^{-}\Delta _{o}^{+}-4G^{2}} , \label{A12}
\\
\Delta _{j}^{\pm } &=&i(\omega \pm \Delta _{j})-\kappa _{j},j=w,o
\label{A16} \\
\Delta _{m}^{\pm } &=&i(\omega \pm \omega _{m})-\gamma _{m} , \label{A18} 
\end{eqnarray}
\begin{eqnarray}
u &=&-\frac{2i\omega _{m}[2G(g_{o}^{\prime 2}e^{-\mathrm{i}\theta
}-g_{o}^{\prime \ast 2}e^{\mathrm{i}\theta })+2i\Delta _{\mathrm{o}%
}|g_{o}^{\prime }|^{2}]}{\Delta _{o}^{-}\Delta _{o}^{+}-4G^{2}}-\frac{4\omega _{m}\Delta _{\mathrm{w}}|g_{\mathrm{w}}^{\prime }|^{2}}{%
\Delta _{\mathrm{w}}^{-}\Delta _{\mathrm{w}}^{+}}+\Delta _{m}^{-}\Delta
_{m}^{+}. \notag\label{A19} \\  
\end{eqnarray}%
}

\section{The calculation of Signal-to-noise ratio}

\label{The calculation of Signal-to-noise} In order to calculate Eq. (\ref{P}%
) and (\ref{SNR}), firstly we calculate $\langle N_{\eta ,\pm }\rangle
_{H_{j}}$ as follows {\small 
\begin{eqnarray}
\langle N_{\eta ,\pm }\rangle _{H_{1}} &=&\frac{1}{2}(D_{1}\bar{n}%
_{w}^{T}+D_{2}\bar{n}_{o}^{T}+D_{3}\bar{n}_{b}^{T}+D_{4}\bar{n}%
_{B}^{T}+F_{1}+F_{2}+F_{3}+F_{4}),  \label{photoncounts1} \\
\langle N_{\eta ,\pm }\rangle _{H_{0}} &=&\langle N_{\eta ,\pm }\rangle
_{H_{1}}(\eta \rightarrow 0),
\end{eqnarray}%
} where the coefficients of Eq.(\ref{photoncounts1}) is {\small 
\begin{eqnarray}
D_{1} &=&[\sqrt{\eta }B_{2}^{\ast }(\omega )A_{1}(-\omega )+\sqrt{\eta }%
B_{1}^{\ast }(\omega )A_{2}^{\ast }(\omega )\pm B_{2}^{\ast }(\omega )] 
\notag \\
&&\times \lbrack \sqrt{\eta }B_{1}(\omega )A_{2}(\omega )+\sqrt{\eta }%
B_{2}(\omega )A_{1}^{\ast }(-\omega )\pm B_{2}(\omega )] \\
&&+[\sqrt{\eta }B_{2}^{\ast }(\omega )A_{2}(-\omega )+\sqrt{\eta }%
B_{1}^{\ast }(\omega )A_{1}^{\ast }(\omega )\pm B_{1}^{\ast }(\omega )] 
\notag \\
&&\times \lbrack \sqrt{\eta }B_{1}(\omega )A_{1}(\omega )+\sqrt{\eta }%
B_{2}(\omega )A_{2}^{\ast }(-\omega )\pm B_{1}(\omega )],  \notag \\
&&  \notag \\
D_{2} &=&[\sqrt{\eta }B_{2}^{\ast }(\omega )A_{3}(-\omega )+\sqrt{\eta }%
B_{1}^{\ast }(\omega )A_{4}^{\ast }(\omega )\pm B_{4}^{\ast }(\omega )] 
\notag \\
&&\times \lbrack \sqrt{\eta }B_{1}(\omega )A_{4}(\omega )+\sqrt{\eta }%
B_{2}(\omega )A_{3}^{\ast }(-\omega )\pm B_{4}(\omega )) \\
&&+[\sqrt{\eta }B_{2}^{\ast }(\omega )A_{4}(-\omega )+\sqrt{\eta }%
B_{1}^{\ast }(\omega )A_{3}^{\ast }(\omega )\pm B_{3}^{\ast }(\omega )] 
\notag \\
&&\times \lbrack \sqrt{\eta }B_{1}(\omega )A_{3}(\omega )+\sqrt{\eta }%
B_{2}(\omega )A_{4}^{\ast }(-\omega )\pm B_{3}(\omega )]  \notag \\
&&+B_{4}^{\ast }(\omega )B_{4}(\omega )+B_{3}^{\ast }(\omega )B_{3}(\omega ),
\notag \\
&&  \notag \\
D_{3} &=&[\sqrt{\eta }B_{2}^{\ast }(\omega )A_{5}(-\omega )+\sqrt{\eta }%
B_{1}^{\ast }(\omega )A_{6}^{\ast }(\omega )\pm B_{6}^{\ast }(\omega )] 
\notag \\
&&\times \lbrack \sqrt{\eta }B_{1}(\omega )A_{6}(\omega )+\sqrt{\eta }%
B_{2}(\omega )A_{5}^{\ast }(-\omega )\pm B_{6}(\omega )] \\
&&+[\sqrt{\eta }B_{2}^{\ast }(\omega )A_{6}(-\omega )+\sqrt{\eta }%
B_{1}^{\ast }(\omega )A_{5}^{\ast }(\omega )\pm B_{5}^{\ast }(\omega )] 
\notag \\
&&\times \lbrack \sqrt{\eta }B_{1}(\omega )A_{5}(\omega )+\sqrt{\eta }%
B_{2}(\omega )A_{6}^{\ast }(-\omega )\pm B_{5}(\omega ))  \notag \\
&&+B_{6}^{\ast }(\omega )B_{6}(\omega )+B_{5}^{\ast }(\omega )B_{5}(\omega ),
\notag \\
&&  \notag \\
D_{4} &=&(1-\eta )(B_{2}^{\ast }(\omega )B_{2}(\omega )+B_{1}^{\ast }(\omega
)B_{1}(\omega )),
\end{eqnarray}
\begin{eqnarray}
F_{1} &=&[\sqrt{\eta }B_{2}^{\ast }(\omega )A_{1}(-\omega )+\sqrt{\eta }%
B_{1}^{\ast }(\omega )A_{2}^{\ast }(\omega )\pm B_{2}^{\ast }(\omega )] 
\notag \\
&&\times \lbrack \sqrt{\eta }B_{1}(\omega )A_{2}(\omega )+\sqrt{\eta }%
B_{2}(\omega )A_{1}^{\ast }(-\omega )\pm B_{2}(\omega )] ,\\
&&  \notag \\
F_{2} &=&[\sqrt{\eta }B_{2}^{\ast }(\omega )A_{3}(-\omega )+\sqrt{\eta }%
B_{1}^{\ast }(\omega )A_{4}^{\ast }(\omega )\pm B_{4}^{\ast }(\omega )] 
\notag \\
&&\times \lbrack \sqrt{\eta }B_{1}(\omega )A_{4}(\omega )+\sqrt{\eta }%
B_{2}(\omega )A_{3}^{\ast }(-\omega )\pm B_{4}(\omega )]  \notag \\
&&+B_{4}^{\ast }(\omega )B_{4}(\omega ) ,\\
&&  \notag \\
F_{3} &=&[\sqrt{\eta }B_{2}^{\ast }(\omega )A_{5}(-\omega )+\sqrt{\eta }%
B_{1}^{\ast }(\omega )A_{6}^{\ast }(\omega )\pm B_{6}^{\ast }(\omega )] \\
&&\times[\sqrt{\eta }B_{1}(\omega )A_{6}(\omega )+\sqrt{\eta }B_{2}(\omega
)A_{5}^{\ast }(-\omega )\pm B_{6}(\omega )]  \notag \\
&&+B_{6}^{\ast }(\omega )B_{6}(-\omega ) , \notag \\
&&  \notag \\
F_{4} &=&(1-\eta )B_{2}^{\ast }(\omega )B_{2}(-\omega ).
\end{eqnarray}%
} 

Then $\langle (\Delta N_{\eta ,+}-\Delta N_{\eta ,-})^{2}\rangle _{H_{j}}$
can be calculated as 
\begin{eqnarray}
\langle (\Delta N_{\eta ,+}-\Delta N_{\eta ,-})^{2}\rangle _{H_{j}}
&=&\langle N_{\eta ,+}\rangle _{H_{j}}(\langle N_{\eta ,+}\rangle
_{H_{j}}+1)+\langle N_{\eta ,-}\rangle _{H_{j}}(\langle N_{\eta ,-}\rangle
_{H_{j}}  \notag \\
&&+1)- \frac{(\langle d_{\eta ,\mathrm{o}}^{\dagger }d_{\eta ,\mathrm{o}%
}\rangle _{H_{j}}-\langle d_{\mathrm{o}}^{\dagger }{d_{\mathrm{o}}\rangle }%
)^{2}}{2},  \label{fluctuation}
\end{eqnarray}
with $\langle d_{\mathrm{o}}^{\dagger }d_{\mathrm{o}}\rangle $ and $\langle
d_{\eta ,\mathrm{o}}^{\dagger }d_{\eta ,\mathrm{o}}\rangle _{H_{j}}$ in Eq.(%
\ref{fluctuation}) being {\small 
\begin{eqnarray}
\langle d_{\mathrm{o}}^{\dagger }d_{\mathrm{o}}\rangle &=&(B_{2}^{\ast
}(\omega )B_{2}(\omega )+B_{1}^{\ast }(\omega )B_{1}(\omega ))\bar{n}_{%
\mathrm{w}}^{T}+(B_{4}^{\ast }(\omega )B_{4}(\omega )+  \notag \\
&&B_{3}^{\ast }(\omega )B_{3}(\omega ))\bar{n}_{\mathrm{o}}^{T}+(B_{6}^{\ast
}(\omega )B_{6}(\omega )+B_{5}^{\ast }(\omega )B_{5}(\omega ))\bar{n}_{%
\mathrm{b}}^{T}+  \notag \\
&&B_{2}^{\ast }(\omega )B_{2}(\omega )+B_{4}^{\ast }(\omega )B_{4}(\omega
)+B_{6}^{\ast }(\omega )B_{6}(\omega ), \\
&&  \notag \\
\langle d_{\eta ,\mathrm{o}}^{\dagger }d_{\eta ,\mathrm{o}}\rangle _{H_{1}}
&=&K_{1}\bar{n}_{\mathrm{w}}^{T}+K_{2}\bar{n}_{\mathrm{o}}^{T}+K_{3}\bar{n}_{%
\mathrm{b}}^{T}+K_{4}\bar{n}_{B}^{T}+\sum_{i=1}^{4}T_{i},  \label{reflected1}
\\
&&  \notag \\
\langle d_{\eta ,\mathrm{o}}^{\dagger }d_{\eta ,\mathrm{o}}\rangle _{H_{0}}
&=&\langle d_{\eta ,\mathrm{o}}^{\dagger }d_{\eta ,\mathrm{o}}\rangle
_{H_{1}}(\eta \rightarrow 0),  \label{reflected0}
\end{eqnarray}%
} where the coefficient of Eq.(\ref{reflected1}) is {\small 
\begin{eqnarray}
K_{1} &=&\eta \lbrack (B_{2}^{\ast }(\omega )A_{1}(-\omega )+B_{1}^{\ast
}(\omega )A_{2}^{\ast }(\omega ))(B_{1}(\omega )A_{2}(\omega )+B_{2}(\omega
)A_{1}^{\ast }(-\omega ))+  \notag \\
&&~~(B_{2}^{\ast }(\omega )A_{2}(-\omega )+B_{1}^{\ast }(\omega )A_{1}^{\ast
}(\omega ))(B_{1}(\omega )A_{1}(\omega )+B_{2}(\omega )A_{2}^{\ast }(-\omega
))],\notag\\
\end{eqnarray}%
\begin{eqnarray}
K_{2} &=&\eta \lbrack (B_{2}^{\ast }(\omega )A_{3}(-\omega )+B_{1}^{\ast
}(\omega )A_{4}^{\ast }(\omega ))(B_{1}(\omega )A_{4}(\omega )+B_{2}(\omega
)A_{3}^{\ast }(-\omega ))+  \notag \\
&&~~(B_{2}^{\ast }(\omega )A_{4}(-\omega )+B_{1}^{\ast }(\omega )A_{3}^{\ast
}(\omega ))(B_{1}(\omega )A_{3}(\omega )+B_{2}(\omega )A_{4}^{\ast }(-\omega
))]+  \notag \\
&&~~B_{4}^{\ast }(\omega )B_{4}(\omega )+B_{3}^{\ast }(\omega )B_{3}(\omega ),
\end{eqnarray}%
\begin{eqnarray}
K_{3} &=&\eta \lbrack (B_{2}^{\ast }(\omega )A_{5}(-\omega )+B_{1}^{\ast
}(\omega )A_{6}^{\ast }(\omega ))(B_{1}(\omega )A_{6}(\omega )+B_{2}(\omega
)A_{5}^{\ast }(-\omega ))+  \notag \\
&&~~(B_{2}^{\ast }(\omega )A_{6}(-\omega )+B_{1}^{\ast }(\omega )A_{5}^{\ast
}(\omega ))(B_{1}(\omega )A_{5}(\omega )+B_{2}(\omega )A_{6}^{\ast }(-\omega
))]+  \notag \\
&&~~B_{6}^{\ast }(\omega )B_{6}(\omega )+B_{5}^{\ast }(\omega )B_{5}(\omega ),
\\
K_{4} &=&(1-\eta )[B_{2}^{\ast }(\omega )B_{2}(\omega )+B_{1}^{\ast }(\omega
)B_{1}(\omega )],
\end{eqnarray}%
\begin{eqnarray}
T_{1} &=&\eta \lbrack (B_{2}^{\ast }(\omega )A_{1}(-\omega )+B_{1}^{\ast
}(\omega )A_{2}^{\ast }(\omega ))(B_{1}(\omega )A_{2}(\omega )+B_{2}(\omega
)A_{1}^{\ast }(-\omega ))],  \notag \\
&& \\
T_{2} &=&\eta \lbrack (B_{2}^{\ast }(\omega )A_{3}(-\omega )+B_{1}^{\ast
}(\omega )A_{4}^{\ast }(\omega ))(B_{1}(\omega )A_{4}(\omega )+B_{2}(\omega
)A_{3}^{\ast }(-\omega ))]  \notag \\
&&+B_{4}^{\ast }(\omega )B_{4}(\omega ), \\
&&  \notag \\
T_{3} &=&\eta \lbrack (B_{2}^{\ast }(\omega )A_{5}(-\omega )+B_{1}^{\ast
}(\omega )A_{6}^{\ast }(\omega ))(B_{1}(\omega )A_{6}(\omega )+B_{2}(\omega
)A_{5}^{\ast }(-\omega ))]  \notag \\
&&+B_{6}^{\ast }(\omega )B_{6}(\omega ), \\
&&  \notag \\
T_{4} &=&(1-\eta )B_{2}^{\ast }(\omega )B_{2}(\omega ).  \label{photoncounts2}
\end{eqnarray}%
}


$\ast$ and $*$

\end{document}